\documentstyle[12pt]{article}
\begin{document}
\vspace*{2cm}
\begin{center}
\begin{large}
{\bf Exclusion Statistics of Composite Fermions}
\end{large}
\end{center}

\vspace{1cm}
\begin{center}
Piotr Sitko 
\vspace{0.3cm}

  Institute of Physics, 
Wroc\l{}aw University of Technology,\\ Wybrze\.ze Wyspia\'{n}skiego
 27, 50-370 Wroc\l{}aw, Poland.
\end{center}
 
 \vspace{1cm}
\begin{center}

{\bf Abstract}

\end{center}
The  exclusion statistics parameter of composite fermions is determined
as an odd number ($\alpha=3$, $5$, ...).
The statistics of composite fermion excitations at 
$\nu= \frac{n}{2pn+1}$ is rederived as $\alpha_{qe}^{CF}=1+\frac{2p}{2pn+1}$,
$\alpha_{qh}^{CF}=1-\frac{2p}{2pn+1}$.
The duality $\frac{1}{\alpha_{qe}(n,2p)}=\alpha_{qh}(n+1,2p)$
is found.
The distribution function for $\alpha=3$ is obtained.

\vspace{1cm}
\noindent
%PACS: 71.10.Pm, 5.30.Pr\\
%keywords: fractional quantum Hall effect, composite fermions, 
%exclusion statistics

\vspace{0.5cm}
\noindent {\bf Introduction}
\vspace{0.5cm}

The exclusion statistics was proposed by Haldane \cite{Haldane} as
generalization of Pauli exclusion principle.
The idea of Haldane is to define the change in the available particle
space when particles are added to the system (or removed):
\begin{equation}
\Delta d_{i}=-\sum_{j}\alpha_{ij} \Delta N_{j}\; .
\end{equation}
The generalization of the Haldane exclusion statistics was proposed by Wu
\cite{Wu} and others \cite{Nayak,Rajagopal,Ouvry}.
The idea of generalization is to divide the available particle space
for smaller cells (let us  say with $k$ states). 
The number of many-particle states
is given by:
\begin{equation}
\label{def}
\prod_{i}  {d_{N_{i}}+N_{i}-1 \choose N_{i}}
\end{equation}
where  $d_{N_{i}}=k-\alpha (N_{i}-1)$, $\alpha=\alpha_{ii}$.

Johnson and Canright used the spherical geometry to find the statistics
of Laughlin quasiparticles \cite{Canright}. Analyzing numerical results
they found the number of  many-particle states of
quasiholes (of the Laughlin $1/m$ state, $m$ -- odd number):
\begin{equation}
\label{Canright}
N_{e}+N_{qh} \choose N_{qh}
\end{equation}
where $N_{qh}$ is the number of quasiholes,
$2S+1=m(N_{e}-1)+1-N_{qh}$, $2S$ is the number of flux quanta piercing the
sphere. According to the definition (\ref{def})  $\alpha_{qh}=\frac{1}{m}$
\cite{Canright}.  The corresponding statistics parameter of Laughlin
 quasielectrons is  $\alpha_{qe}=2-\frac{1}{m}$.
Here, we perform similar analysis in order to determine the exclusion
statistics of composite fermions.

\vspace{0.5cm}
\noindent {\bf Composite fermions}
\vspace{0.5cm}

The exact diagonalization results for the sphere can  be interpreted
in terms of composite fermions \cite{Quinn} if the effective field
$2S^{*}=2S-2p(N_{e}-1)$ is introduced.
The composite fermion approach predicts the same angular momentum shell
for quasiparticles \cite{Quinn}, however,
the main role  play  composite fermions (the number of composite fermions
equals the number of
electrons $N_{CF}=N_{e}$).
The Eq. (\ref{Canright}) is:
\begin{equation}
2S^{*}+1 \choose N_{qh}
\end{equation}
and equals
\begin{displaymath}
{N_{CF}+2S -m(N_{CF}-1) \choose N_{CF} }
\end{displaymath}
\begin{equation}
= {2S+1 +(1-m)(N_{CF}-1) \choose N_{CF}}\; .
\end{equation}
According to the definition (\ref{def})
the statistics parameter of composite fermions  $\alpha_{CF}=m$ 
(odd number $\alpha_{CF}=3$, $5$, ... ,
for $\alpha=1$ one has fermions).

Wu {\it et al.} \cite{Wu} found the distribution function of
$\alpha$-particles.
The duality between $\alpha$-particles of $\alpha=1/m$ 
(holes) and $\alpha=m$ was noticed by Nayak and
Wilczek \cite{Nayak}. Our analysis reflects the duality between Laughlin
quasiholes and composite fermions.

The distribution function  is given by the set of equations \cite{Wu}:
\begin{equation}
n_{i}=\frac{1}{w+\alpha}
\end{equation}
\begin{equation}
w^{\alpha}(1+w)^{1-\alpha}=\xi
\end{equation}
\begin{equation}
\xi = e^{\frac{\epsilon_{i}-\mu}{kT}}\; .
\end{equation}
For example, for $\alpha=3$ the solution is
\begin{equation}
w=\frac{1}{s_{+}+s_{-}-\frac{2}{3}}
\end{equation}
where
\begin{equation}
s_{\pm}=[\frac{1}{2\xi}+\frac{1}{27}\pm \sqrt{\frac{1}{\xi}
(\frac{1}{\xi}+\frac{1}{27})}]^{\frac{1}{3}}\; .
\end{equation}

It is interesting to consider also the composite fermion excitations.
Let us consider the filling $\nu=\frac{n}{2pn+1}$, then one gets $n$ filled
shells (in the field $2S^{*}=2S-2p(N_{e}-1)$ 
the degeneration of the $n-th$ effective shell is $2S^{*}+2n-1$). 
When one creates $N_{qe}$ quasielectrons (of the state
$\nu=\frac{n}{2pn+1}$) in the $(n+1)$-th level, the number of many-particle states is
\begin{equation}
2S^{*}_{qe}+2n+1 \choose N_{qe}
\end{equation}
and
\begin{equation}
2S^{*}_{qe}=-\frac{2p}{1+2pn}N_{qe}+\frac{2S-2pn^{2}+2p}{1+2pn}\; .
\end{equation}
If quasiholes are present in the $n$-th level then
\begin{equation}
2S^{*}_{qh}+2n-1 \choose N_{qh}
\end{equation}
and
\begin{equation}
2S^{*}_{qh}=\frac{2p}{1+2pn}N_{qh}+\frac{2S-2pn^{2}+2p}{1+2pn}\; .
\end{equation}
Hence, according to the definition (\ref{def})
\begin{equation}
\label{a1}
\alpha_{qe}=1+\frac{2p}{1+2pn}    \; ,
\end{equation}
\begin{equation}
\label{a2}
\alpha_{qh}=1-\frac{2p}{1+2pn}
\end{equation}
as was first  obtained in Ref. \cite{Johnson2}.
For $n=1$ one gets the Laughlin states $\frac{1}{m}=\frac{1}{2p+1}$.
One can notice the duality between 
quasielectrons of the state $\frac{n}{2pn+1}$
and quasiholes of the state $\frac{n+1}{2p(n+1)+1}$: 
\begin{equation}
\frac{1}{\alpha_{qe}(n,2p)}=\alpha_{qh}(n+1,2p)\; .
\end{equation}
For example, consider quasielectrons at $1/3$ and quasiholes at $2/5$.

\vspace{0.5cm}
\noindent    {\bf Conclusions}
\vspace{0.5cm}

It is found that composite fermions can be described
within the generalized  exclusion statistics
as particles with statistics parameter $\alpha$ being an odd number.
Also, the exclusion statistics of composite fermion excitations is
rederived as  $\alpha_{qe}=1+\frac{2p}{1+2pn},\;
\alpha_{qh}=1-\frac{2p}{1+2pn}$.
The duality $\frac{1}{\alpha_{qe}(n,2p)}=\alpha_{qh}(n+1,2p)$
is found.

This work was supported by KBN grant No. 2 P03B 111 18.

%\newpage
      
%\vspace{1cm}

\end{document}